\documentclass[12pt]{article}

\usepackage{graphics}

\def\eq#1{Eq.~(\ref{#1})}
\newcommand{\secn}[1]{Section~\ref{#1}}
\newcommand{\bra}[1]{\langle{#1}|}
\newcommand{\ket}[1]{|{#1}\rangle}

\def\beq{\begin{equation}}
\def\eeq{\end{equation}}
\def\beqa{\begin{eqnarray}}
\def\eeqa{\end{eqnarray}}
\def\ifm{\ifmmode}
\def\msb{\ifm \overline{\rm MS}\,\, \else $\overline{\rm MS}\,\, $\fi}
\def\msbns{\ifm \overline{\rm MS}\, \else $\overline{\rm MS}\, $\fi}

\newcommand{\sect}[1]{\setcounter{equation}{0}\section{#1}}

\renewcommand{\a}{\alpha}

\renewcommand{\b}{\beta}

\newcommand{\e}{\epsilon}

\newcommand{\hepph}[1]{{\tt hep-ph/#1}}
\newcommand{\NP}[1]{Nucl.\ Phys.\ {\bf #1}}
\newcommand{\PL}[1]{Phys.\ Lett.\ {\bf #1}}

\newcommand{\PR}[1]{Phys.\ Rev.\ {\bf #1}}
\newcommand{\PRL}[1]{Phys.\ Rev.\ Lett.\ {\bf #1}}

\renewcommand{\thefootnote}{\fnsymbol{footnote}}
\def\gtil{{\widetilde {G}}}

\def\bet{\begin{tabular}}
\def\eet{\end{tabular}}

\begin{document}

\begin{titlepage}

\rightline{DFTT-26/2005}
\rightline{NIKHEF/2005--015}

\rightline{\hfill August 2005}

\vspace{1.8cm}

\centerline{\Large \bf Threshold resummation for electroweak} 
\centerline{\Large \bf annihilation from DIS data}

\vskip 1cm

\centerline{\bf Eric Laenen\footnote{e-mail: {\tt t45@nikhef.nl}}}
\centerline{\sl NIKHEF Theory Group} 
\centerline{\sl Kruislaan 409, 1098 SJ Amsterdam, The Netherlands,}
\centerline{\sl and Institute for Theoretical Physics, Utrecht University}
\centerline{\sl Leuvenlaan 4, 3584 CE Utrecht, The Netherlands}

\vspace{2mm}

\centerline{\bf Lorenzo Magnea\footnote{e-mail: {\tt magnea@to.infn.it}}}
\centerline{\sl Dipartimento di Fisica Teorica, Universit\`a di Torino}
\centerline{\sl and INFN, Sezione di Torino}
\centerline{\sl Via P. Giuria 1, I--10125 Torino, Italy}

\vskip 1.2cm
 
\begin{abstract}

We show that higher-order coefficients required to perform threshold
resummation for electroweak annihilation processes, such as Drell-Yan
or Higgs production via gluon fusion, can be computed using
perturbative results derived in Deep Inelastic Scattering. As an
example, we compute the three-loop coefficient $D^{(3)}$, generating
most of the fourth tower of threshold logarithms for the Drell-Yan
cross section in the \msb scheme, using the recent three-loop results
for splitting functions and for the quark form factor, as well as a
class of exponentiating two-loop contributions to the Drell-Yan
process.

\end{abstract}

\end{titlepage}

\newpage

\renewcommand{\thefootnote}{\arabic{footnote}}
\setcounter{footnote}{0}
\setcounter{page}{1}

\sect{Introduction}
\label{intro}

Soft gluon resummations
\cite{Sterman:1987aj,Catani:1989ne,Forte:2002ni} have proven to be a
valuable tool in perturbative QCD.  They have provided a deep
understanding of the structure of perturbation theory to all orders,
which has in turn opened the door to studies on nonperturbative
effects, and they have also been extensively used in phenomenology,
broadening the range of QCD predictions towards the edges of phase
space, where even hard processes are dominated by multiple soft gluon
radiation.

Resummation is closely related to factorization
\cite{Contopanagos:1997nh}. For threshold resummations, the hard
partonic cross section for a given QCD process can be expressed as a
convolution (with respect to the energy fraction carried by hard
partons, $x$) of different functions responsible for soft, collinear
and hard radiation. The convolution turns into an ordinary product
upon taking a Mellin transform. Logarithmic enhancements as $x \to 1$
turn into logarithms of the Mellin variable $N$, and these logarithms
can be shown to exponentiate, using evolution equations for the
various functions involved in the factorization.

To be precise, the resummed exponent is expressed in terms of moments
of distributions singular as $x \to 1$,
\beq
 {\cal D}_k (N) \equiv \int_0^1 x^{N - 1} \left( 
 \frac{\log^k (1 - x)}{1 - x} \right)_+ = \frac{(- 1)^{k + 1}}{k + 1} 
 \log^{k + 1} N + {\cal O} \left(\log^k N \right)~,
\label{calD}
\eeq
as well as terms independent of $N$, corresponding to moments of
$\delta (1 - x)$~\cite{Eynck:2003fn}. The pattern of exponentiation is
nontrivial: in general, a perturbative calculation will contain terms
of the form $\alpha_s^k \log^{2 k} N$ multiplying the Born cross
section, whereas in the exponent one finds at most terms of the form
$\alpha_s^k \log^{k + 1} N$. Furthermore, a $g$-loop resummed
calculation will determine completely the coefficients of the terms in
the exponent proportional to $\alpha_s^k \log^{k + 2 - g} N$, to all
orders in $\alpha_s$. Such terms are usually described as ${\rm N}^{g
- 1}{\rm LL}$, with leading logarithms (LL) determined at one loop,
next-to-leading logarithms (NLL) determined at two loops, and so
forth.

Recently, the scope and expected precision of a range of QCD
calculations have been extended in a remarkable series of papers by
Moch, Vermaseren and Vogt (MVV), who computed first the three-loop
contribution to the QCD splitting
functions~\cite{Moch:2004pa,Vogt:2004mw}, and then the complete
three-loop DIS coefficient functions~\cite{Vermaseren:2005qc}, in what
is arguably the most complex perturbative calculation ever carried out
in quantum field theory. Their results both test and extend the range
of threshold resummation for DIS, which can now be performed exactly
to ${\rm N}^2{\rm LL}$ accuracy. Furthermore, ${\rm N}^3{\rm LL}$
terms can also be determined, up to a single unknown coefficient
requiring a four-loop calculation, the fourth-order contribution to
the cusp anomalous dimension of a Wilson line in the $\msb$ scheme. It
can, however, be argued convincingly that the numerical effect of this
coefficient is negligible~\cite{Moch:2005ba}. Thus soft resummation
for DIS can now be tested at the level of the fourth tower of
logarithms, providing nontrivial checks on the convergence of the
expansion as the logarithmic accuracy is increased.

Another class of benchmark cross sections for soft gluon resummation
is given by electroweak annihilation processes in hadronic collisions,
comprising Drell-Yan dimuon production, electroweak boson production,
and Higgs production via gluon fusion. The inclusive cross sections
for these processes are known to
NNLO~\cite{Hamberg:1991np,vanNeerven:1992gh,Harlander:2002wh}, and
with the knowledge of the three-loop splitting functions the
corresponding resummation can now be performed exactly at ${\rm
N}^2{\rm LL}$ level, both in the $\msb$ and in the DIS factorization
schemes. Lacking a three-loop calculation, however, ${\rm N}^3{\rm
LL}$ terms are unknown, except for running coupling effects. It is the
purpose of this letter to show that, using only results extracted from
the three-loop DIS calculations of MVV, as well as known two-loop
perturbative results for electroweak annihilation, one can bring the
accuracy of threshold resummation for these processes in line with
DIS, performing ${\rm N}^3{\rm LL}$ resummation up to the unknown, and
very likely negligible, contribution of the four-loop cusp anomalous
dimension.

In the following, we will concentrate on the Drell-Yan cross section
in the $\msb$ factorization scheme, although the reasoning is readily
generalized to other electroweak annihilation processes and to the DIS
scheme. We will make use of a factorization derived
in~\cite{Eynck:2003fn}, where the complete exponentiation of
$N$-independent terms was proven, to show that the coefficients of
single-logarithmic contributions at $g$ loops in the resummed exponent
are completely determined by the knowledge of the $g$-loop nonsinglet
splitting function, simple poles in the $g$-loop quark form factor,
and $N$-independent terms at $g - 1$ loops in the Drell-Yan cross
section. We will explicitly compute these coefficients at the
three-loop level, and provide a general ansatz for their expression to
all orders. These results will be useful in refining the theoretical
prediction for processes of great interest at the LHC, such as $Z_0$
production and Higgs production via gluon fusion, by extending our
knowledge of soft-gluon effects, and our control of the theoretical
uncertainty due to uncalculated higher-order perturbative as well as
nonperturbative corrections.

\sect{Factorization and exponentiation}
\label{expon}

Our starting point is the unsubtracted partonic cross section for the
Drell-Yan process. Near partonic threshold, its Mellin moments can be
factorized as \cite{Sterman:1987aj,Eynck:2003fn}
\beq
 \omega (N, \e) = |\Gamma(Q^2,\e)|^2 \, \left( \psi_R(N, \e) \right)^2 \,
 U_R(N, \e) + {\cal O}(1/N)~.
\label{om2}
\eeq
Here $\psi_R(N, \e)$ is the Mellin transform of a quark distribution
at defined energy fraction, responsible for collinear divergences,
$U_R(N, \e)$ is an eikonal function describing the effects of soft
gluon radiation at large angles, and $\Gamma(Q^2,\e)$ is the
(timelike) quark form factor.  Near threshold, where all gluon
radiation is soft, the quark distribution obeys a Sudakov-type
evolution equation which can be solved in exponential form, as
\beq
\label{epsir}
 \psi_R(N, \e) = \exp \left\{ \int_0^1 dz ~\frac{z^{N - 1}}{1 - z} 
 \int_z^1 \frac{dy}{1 - y} ~\kappa_\psi \left(\overline{\a} 
 \left((1 - y)^2 Q^2 \right), \e \right) \right\}\,.
\eeq
Similarly, eikonal exponentiation applies to the soft function $U_R$, which 
can be written as
\beq
\label{eur}
 U_R(N, \e) = \exp\left\{- \int_0^1 dz ~\frac{z^{N - 1}}{1 - z} \,
 g_U \left( \overline{\a} \left((1 - z)^2 Q^2\right), \e 
 \right) \right\}\,.
\eeq
The electromagnetic quark form factor $\Gamma$, on the other hand, is 
defined by
\beq 
 \Gamma_\mu (p_1, p_2; \mu^2, \e) \equiv \bra{0}
 J_\mu (0) \ket{{p_1,p_2}} \label{def} = - {\rm i} e\, e_q ~
 \overline{v}(p_2) \gamma_\mu u(p_1) ~\Gamma 
 \left( Q^2, \e \right)~, \nonumber
\label{ffdef}
\eeq
and it is one of the best understood amplitudes in perturbative
QCD. Its logarithmic dependence on the scale $Q^2$ can be determined
using renormalization group and gauge
invariance~\cite{Mueller:1979ih,Collins:1980ih,Sen:1981sd}, and the
resulting evolution equation can be solved explicitly in dimensional
regularization~\cite{Magnea:1990zb}, yielding the exponential
expression
\beq
\label{mygam}
 \Gamma(Q^2, \e) = \exp \Bigg\{
 \frac{1}{2} \int_0^{- Q^2} \frac{d\xi^2}{\xi^2} \Bigg[
 K \left(\a_s, \e \right) +
 G \left(\overline{\a} \left(\xi^2 \right), \e \right) + 
 \frac{1}{2} \int_{\xi^2}^{\mu^2} \frac{d \lambda^2}{\lambda^2}
 \gamma_K \left( \overline{\a} \left(\lambda^2\right) \right) 
 \Bigg] \Bigg\}~,
\eeq
where $\gamma_K (\alpha_s)$ is the cusp anomalous
dimension~\cite{Korchemsky:1987wg,Korchemsky:1988hd}, $G(\alpha_s,
\e)$ collects all other scale-dependent terms, and is finite as $\e
\to 0$, while $K(\alpha_s, \e)$ is a pure counterterm.  A key feature
of Eqs.~(\ref{epsir})--(\ref{mygam}) is the usage of the $d$-dimensional
running coupling $\overline{\alpha} (\xi^2)$, defined in $d = 4 - 2
\e$ by the equation
\beq
 \xi \frac{\partial \overline{\alpha}}{\partial \xi} \equiv 
 \b (\e, \overline{\alpha}) = - 2 \e 
 \overline{\alpha} + \hat{\b} (\overline{\alpha}) ~~,~~~
 \hat{\b} (\overline{\alpha}) = - \frac{\overline{\alpha}^2}{2
 \pi} \sum_{n = 0}^\infty b_n \left( 
 \frac{\overline{\alpha}}{\pi} \right)^n~,
\label{dalpha}
\eeq
where $b_0 = (11 C_A - 2 n_f)/3$ and $b_1 = (17 C_A^2 - 5 C_A n_f - 3
C_F n_f)/6$ in our normalization. Through $\overline{\alpha}$,
integration over the scale of the coupling generates {\it all}
infrared and collinear poles in Eqs.~(\ref{epsir})--(\ref{mygam}), so
that all functions appearing in the exponents are finite as $\e \to
0$, with the exception of the counterterm $K$ in the quark form
factor, whose only effect however is to cancel singularities arising
from the $\xi$-independent limit of integration in the integral of the
anomalous dimension $\gamma_K$. Further, dimensional continuation of
the coupling regulates the Landau pole, which lies on the integration
contour in $d = 4$, allowing for an explicit evaluation of the
exponents in terms of analytic functions of $\alpha_s$ and
$\e$~\cite{Magnea:2000ss,Magnea:2001ge}.

Our next task is to perform mass factorization on \eq{om2}. We do it
here in the $\msb$ scheme, where we can make use of the
expression~\cite{Contopanagos:1997nh}
\beq
\label{phims}
 \phi_{\msb}(N, \e) = \exp\Bigg[
 \int_0^{Q^2}  \frac{d \xi^2}{\xi^2} \, \Bigg\{
 \int_0^1 \! dz \, \frac{z^{N - 1} - 1}{1 - z}
 A \left( \overline{\a} \left(\xi^2 \right) \right)
 + B_\delta \left(\overline{\a} \left(\xi^2 \right) 
 \right) \Bigg\} \Bigg] + {\cal O} \left(\frac{1}{N} \right)\,.
\eeq
Here $A(\alpha_s)$ can be extracted from the singular behavior of the
nonsinglet QCD splitting functions as $z \to 1$, and is known to be
related to the cusp anomalous dimension by $A(\alpha_s) =
\gamma_K(\alpha_s)/2$, while $B_\delta (\alpha_s)$ is the coefficient
of $\delta(1 - x)$ in the same splitting function. Once again, it is
easy to see that $\phi_{\msb}(N, \e)$ is a pure counterterm, with all
poles generated by integration over the running coupling. Clearly,
\eq{phims} is a simple exponentiation of the splitting function in the
IR limit, including running coupling effects. Since it does not have
an obvious diagrammatic interpretation (see, however,
Ref.~\cite{Berger:2002sv}), there is a certain amount of arbitariness
in distinguishing real and virtual contributions in \eq{phims}. This
arbitrariness was exploited in Ref.~\cite{Eynck:2003fn} to define
\beq
\label{phivr}
 \phi_{\msb}(N, \e) =  \phi_V(\e) \: 
 \phi_R(N, \e)\,,
\eeq
where
\beq
\label{phiv}
 \phi_{V} (\e) = \exp \Bigg\{
 \frac{1}{2} \int_0^{Q^2} \frac{d\xi^2}{\xi^2} \Bigg[
 K \left(\a_s, \e \right) +
 \gtil \left(\overline{\a} \left(\xi^2\right) \right) +
 \frac{1}{2} \int_{\xi^2}^{\mu^2} \frac{d \lambda^2}{\lambda^2}
 \gamma_K \left( \overline{\a} \left(\lambda^2 \right) 
 \right) \Bigg] \Bigg\}~.
\eeq
The structure of \eq{phiv} clearly mimicks that of the quark form
factor, \eq{mygam}, and in fact it is designed so that $\phi_{V} (\e)$
will precisely cancel all IR and collinear poles arising from
$\Gamma(Q^2, \e)$. This requirement, together with the requirement
that $\phi_{V} (\e)$ be a pure counterterm, uniquely fixes the new
function $\gtil (\alpha_s)$, which can be determined recursively from
$G(\alpha_s, \e)$, as was done explicitly in
Ref.~\cite{Eynck:2003fn}. We are now ready to give our final
expression for the Drell-Yan partonic cross section in the $\msb$
scheme, which is
\beq
\label{omms}
 \widehat{\omega}_{\msb}(N) \equiv \frac{\omega(N,
 \e)}{\left( \phi_{\msb}(N, \e) \right)^2} = 
 \left(\frac{|\Gamma(Q^2, \e)|^2}{\phi_V(\e)^2} \right) \;
 \left[\frac{\left(\psi_R(N, \e)\right)^2 \, U_R(N, \e)}{\left(
 \phi_R(N,\e)\right)^2} \right] + {\cal O}  \left(\frac{1}{N} 
 \right) \, .
\eeq
This expression has the important feature that virtual and real
contributions are separately finite. Factoring out the virtual part
$\widehat{\omega}_{\msb}^{(V)}(N) \equiv |\Gamma(Q^2,
\e)|^2/(\phi_V(\e))^2$, and mapping the real terms to the conventional
expression for the resummed Drell-Yan cross section in the $\msb$
scheme, including $N$-independent terms as done in
Ref.~\cite{Eynck:2003fn}, we are lead to our basic equation
\beqa
 \widehat{\omega}_{\msb}^{(R)} (N) & \equiv & 
 \lim_{\e \to 0} \left[ \frac{\left(\psi_R(N, \e)\right)^2 
 \, U_R(N, \e)}{\left(\phi_R(N, \e)\right)^2} \right] \nonumber \\
 & = & \exp \Bigg[F_{\msbns} (\a_s) + \int_0^1 \! dz \,
 \frac{z^{N - 1} - 1}{1 - z} \Bigg\{ 2 \,\int_{Q^2}^{(1 - z)^2 Q^2} 
 \frac{d \mu^2}{\mu^2} \, A \left(\a_s(\mu^2) \right) \nonumber \\
 & + & D \left(\a_s \left((1 - z)^2 Q^2 \right) \right) \Bigg\} 
 \Bigg] + {\cal O}(1/N)~.
\label{finms}
\eeqa
Eq.~(\ref{finms}) spells out our basic strategy to determine the
resummation coefficients: $\widehat{\omega}_{\msb}^{(R)} (N)$ must be
finite by the factorization theorem, given our construction of the
virtual part $\widehat{\omega}_{\msb}^{(V)} (N)$; the poles arising
from the denominator, furthermore, are completely determined by the
splitting functions and by the quark form factor, as seen from
Eqs.~(\ref{phims}) and~(\ref{phiv}); requiring their cancellation
determines a subset of the perturbative coefficients of the numerator
functions, which are sufficient to control the expansion of the
functions $A$ and $D$.

\sect{Constraints from finiteness}
\label{const}

The scale dependence of $\widehat{\omega}_{\msb}^{(R)} (N)$ can be
explicitly computed order by order making use of the exponential
expressions for the functions $\psi_R$, $U_R$ and $\phi_R$. An
important point is the fact that $\psi_R$ and $U_R$ are
renormalization group invariant~\cite{Sterman:1987aj}, which
determines explicitly the scale dependence of their exponents.
Consider for example the quark distribution $\psi_R$. Imposing RG
invariance leads to
\beq
 \left(\mu \frac{\partial}{\partial \mu} + \beta \left(\e, 
 \a_s \right) \frac{\partial}{\partial \a_s} \right) 
 \kappa_\psi \left(\frac{(1 - y) Q}{\mu}, \a_s(\mu^2), \e 
 \right) = 0~,
\label{renk}
\eeq
which can be solved perturbatively using the explicit expression for 
the $\b$ function, \eq{dalpha}, and writing
\beq
 \kappa_\psi \left( \xi, \a_s, \e \right) = \sum_{n = 1}^\infty
 \left(\frac{\a_s}{\pi}\right)^n \kappa_\psi^{(n)} \left( \xi, \e 
 \right)~,
\label{pexp}
\eeq
where from now on $\xi$ will denote the ratio of the relevant scale
(here $(1 - x) Q$) to the renormalization scale, for which we take
$\mu = Q$. Alternatively, one can impose
\beq
 \kappa_\psi \left( \xi, \a_s, \e \right) = 
 \kappa_\psi \left( 1, \overline{\alpha} (\xi), \e \right) =
 \sum_{n = 1}^\infty
 \left(\frac{\overline{\alpha} (\xi)}{\pi}\right)^n 
 \kappa_\psi^{(n)} \left( 1, \e \right)~,
\label{userun}
\eeq
which also determines the scale dependence of the perturbative
coefficients $\kappa_\psi^{(n)} \left( \xi, \e \right)$. Using for the
running coupling the solution of \eq{dalpha} expanded to three loops
\beqa
 \overline{\a} \left( \xi^2, \alpha_s, \e \right) & = &
 \alpha_s \, \xi^{- 2 \e} +
 \alpha_s^2  \, \xi^{- 4 \e} \, \frac{b_0}{4 \pi \e} 
 \left( 1 - \xi^{2 \e} \right) \\
 &&  \hspace{-2cm} + \,\, \alpha_s^3 \, \xi^{- 6 \e} \, 
 \frac{1}{8 \pi^2 \e} \left[ \frac{b_0^2}{2 \e}
 \left( 1 - \xi^{2 \e} \right)^2 +
 b_1 \left( 1 - \xi^{4 \e} \right) 
 \right]~, \nonumber
\label{alpha3}
\eeqa
one finds
\beqa
 \kappa_\psi^{(1)} \left( \xi, \e \right) & = & \kappa_\psi^{(1)} 
 \left(1, \e \right) \xi^{- 2 \e}~, \nonumber \\
 \kappa_\psi^{(2)} \left( \xi, \e \right) & = & \kappa_\psi^{(2)} 
 \left(1, \e \right) \xi^{- 4 \e} + \frac{b_0}{4 \e}
 \kappa_\psi^{(1)} \left(1, \e \right) \xi^{- 2 \e}
 \left(\xi^{- 2 \e} - 1 \right)~, \\
 \kappa_\psi^{(3)} \left( \xi, \e \right) & = & \kappa_\psi^{(3)} 
 \left(1, \e \right) \xi^{- 6 \e} + \frac{b_0}{2 \e} \left( 
 \kappa_\psi^{(2)} \left(1, \e \right) + \frac{b_0}{4 \e} \kappa_\psi^{(1)} 
 \left(1, \e \right) \right) \xi^{- 4 \e}
 \left(\xi^{- 2 \e} - 1 \right) \nonumber \\
 && - \frac{1}{8 \e} \kappa_\psi^{(1)} \left(1, \e \right)
 \left(\frac{b_0^2}{2 \e} - b_1 \right) \xi^{- 2 \e} 
 \left(\xi^{- 4 \e} - 1 \right)~,
\label{expk}
\eeqa
with analogous results holding for the function $g_U (\xi, \alpha_s,
\e)$.  The last formal step is to use the finiteness of $\kappa_\psi$
and $g_U$ as $\e \to 0$ to expand the $\e$-dependent coefficients as
\beq
 \kappa_\psi^{(p)} (1, \e) = \sum_{k = 0}^\infty \kappa_{\psi,k}^{(p)} \, 
 \e^k ~~,~~~ g_U^{(p)} (1, \e) = \sum_{k = 0}^\infty g_{U,k}^{(p)} \, \e^k~,
\label{epsexp}
\eeq
as well as
\beq
 G(\alpha_s, \e) = \sum_{p = 0}^\infty G^{(p)} (\e) \left( 
 \frac{\alpha_s}{\pi} \right)^p 
 = \sum_{p = 0}^\infty \left(\frac{\alpha_s}{\pi} \right)^p 
 \sum_{k = 0}^\infty G^{(p)}_k \, \e^k~.
\label{Gepsexp}
\eeq
Expanding, in a similar way, the various other functions involved in
\eq{omms} in powers of $\alpha_s/\pi$, one can easily determine the
structure of IR-collinear poles, by computing simple integrals.

It is instructive to briefly examine the information that can be
extracted at the one-loop level. From \eq{finms} one derives
\beqa
 && \lim_{\e \to 0} \Bigg\{ \frac{1}{2 \e^2} 
 \left(\kappa_{\psi,0}^{(1)} - 
 \gamma_K^{(1)}\right) + \frac{1}{\e} \left[\frac{g_{U,0}^{(1)} + 
 \kappa_{\psi,1}^{(1)}}{2} + 2 B_\delta^{(1)} - \gtil^{(1)} + 
 \left( 2 A^{(1)} - \kappa_{\psi,0}^{(1)} \right) {\cal D}_0 (N) \right] 
 \nonumber \\
 && \hspace{8mm} + \; 2 \, \kappa_{\psi,0}^{(1)} \, {\cal D}_1 (N) - 
 \left(g_{U,0}^{(1)} + 
 \kappa_{\psi,1}^{(1)}\right) {\cal D}_0 (N) + \frac{g_{U,1}^{(1)} + 
 \kappa_{\psi,2}^{(1)}}{2} \Bigg\} \nonumber \\
 && = \; F_{\msb}^{(1)} + D^{(1)} {\cal D}_0 (N) +
 4 A^{(1)} {\cal D}_1 (N)~.
\label{oneloop}
\eeqa
The cancellation of double poles requires, unsurprisingly, that
$\kappa_{\psi,0}^{(1)} = \gamma_K^{(1)}$. Cancellation of single poles
yields two equations, since the coefficient of the distribution ${\cal
D}_0 (N)$ must separately vanish. One finds that $A^{(1)} =
\kappa_{\psi,0}^{(1)}/2 = \gamma_K^{(1)}/2$ (the factor of $2$ being a
matter of historical conventions); further, one finds that a
combination of coefficients of $U_R$ and $\psi_R$ is determined by
$\phi_R$, yielding
\beq
 g_{U,0}^{(1)} + \kappa_{\psi,1}^{(1)} = - 4 B_\delta^{(1)} + 2 \gtil^{(1)}~.
\label{simpol}
\eeq
Turning our attention to finite terms, we see first that the
coefficient of the leading distribution ${\cal D}_1 (N)$ is confirmed
to be $A^{(1)} = \gamma_K^{(1)}/2$: had we not assumed the function
$A(\alpha_S)$ appearing in $\phi_R$ to be the same as the one
featuring in the resummation, this result would now have been derived
at one loop. Next we see that single logarithms are given by {\it the
same} combination of Drell-Yan coefficients that was determined by the
cancellation of simple poles. This determines $D^{(1)}$ in terms of
DIS data as
\beq
 D^{(1)} = 4 B_\delta^{(1)} - 2 \gtil^{(1)}~.
\label{D1}
\eeq
Finally, the one-loop exponentiated constants are given by
$F_{\msb}^{(1)} = (g_{U,1}^{(1)} + \kappa_{\psi,2}^{(1)})/2$.

Clearly, all the coefficients involved at one loop are known or easily
computed. For example one finds~\cite{Sterman:1987aj}, in the $\msb$
scheme,
\beq
 \kappa_\psi^{(1)} \left(1, \e \right) = 2 C_F {\rm e}^{\gamma_E \e}
 \frac{\Gamma(2 - \e)}{\Gamma(2 - 2 \e)}~~,~~~
 g_U^{(1)} \left(1, \e \right) = - 2 C_F {\rm e}^{\gamma_E \e}
 \frac{\Gamma(1 - \e)}{\Gamma(2 - 2 \e)}~,
\label{onlog}
\eeq
while, as derived in~\cite{Eynck:2003fn}, $\gtil^{(1)} = G_0^{(1)} = 3
C_F/2$.  It is well-known that $B_\delta^{(1)} = 3 C_F/4$, so one
finds consistently
\beq
 D^{(1)} = 0 ~~,~~~ F_{\msb}^{(1)} = - \frac{3}{2} \zeta(2) C_F~,
\label{outone}
\eeq
as confirmed by a direct one-loop calculation of the Drell-Yan cross
section.

At two loops, the pattern repeats iself with a few twists. The
cancellation of triple and double poles brings in no new information,
except the fact that the function $\kappa_\psi$ begins to differ from
$\gamma_K$ by running coupling effects,
\beq
 \kappa_{\psi,0}^{(2)} = \gamma_K^{(2)} + \frac{b_0}{2} \left( 
 g_{U,0}^{(1)} + \frac{3}{2} \kappa_{\psi,1}^{(1)}\right) = 
 \gamma_K^{(2)} + \frac{1}{2} b_0 C_F~.
\label{kappa2}
\eeq
This however is just a reshuffling between $\psi_R$ and $U_R$, in fact
at the level of single poles the effect cancels and one finds, as
expected, that requiring the cancellation of ${\cal D}_0(N)/\e$ terms
yields $A^{(2)} = \gamma_K^{(2)}/2$
\cite{Kodaira:1981nh,Korchemsky:1992xv}. $N$-independent single-pole
terms, on the other hand, constrain a combination of coefficients of
$g_U$ and $\kappa_\psi$, namely
\beq
 g_{U,0}^{(2)} + \frac{\kappa_{\psi,1}^{(2)}}{2} = - 4 B_\delta^{(2)} +
 2 \gtil^{(2)} + \frac{b_0}{4} \left(g_{U,1}^{(1)} + \frac{3}{2}
 \kappa_{\psi,2}^{(1)} \right)~.
\label{simpol2}
\eeq
Turning to finite terms, one finds that once again running coupling
effects involving $\psi_R$ and $U_R$ cancel, and single logarithms are
determined by
\beq
 D^{(2)} = 4 B_\delta^{(2)} - 2 \gtil^{(2)} - \frac{b_0}{4}
 \left( g_{U,1}^{(1)} + \kappa_{\psi,2}^{(1)} \right) =
 4 B_\delta^{(2)} - 2 \gtil^{(2)} - \frac{b_0}{2} F_{\msb}^{(1)}~.
\label{D2}
\eeq
All required ingredients are known: $B_\delta^{(2)}$ from
Refs.~\cite{Herrod:1980rm,Curci:1980uw}, while $\gtil^{(2)} =
G^{(2)}_0 - b_0 G^{(1)}_1/4$ was given
in~\cite{Eynck:2003fn}\footnote{Notice however a misprint in Eq. (4.6)
of Ref.~\cite{Eynck:2003fn}: the coefficient of $C_A C_F$ in
$G^{(2)}_0$ should read $(2545/108 + 11 \zeta(2)/3 - 13
\zeta(3)))/4$.}. One finds then
\beq
 D^{(2)} = \left(- \frac{101}{27} + \frac{11}{3} \zeta (2) + \frac{7}{2}
 \zeta (3) \right) C_A C_F + \left(\frac{14}{27} - \frac{2}{3} \zeta (2)
 \right) n_f C_F~,
\label{outD2}
\eeq
which agrees with a direct
comparison~\cite{Contopanagos:1997nh,Vogt:2000ci} with the two-loop
calculation of Ref.~\cite{Hamberg:1991np}, in the spirit
of~\cite{Magnea:1991qg}. Exponentiated two-loop constants are also
constrained by
\beq
 F_{\msbns}^{(2)} = \frac{1}{4} \left(g_{U,1}^{(2)} + 
 \frac{\kappa_{\psi,2}^{(2)}}{2} \right) - \frac{b_0}{16} \left( 
 g_{U,2}^{(1)} + \frac{3}{2} \kappa_{\psi,3}^{(1)} \right)~,
\label{outF2}
\eeq
where running coupling effects are readily evaluated using \eq{onlog}.

\sect{The coefficients $D^{(k)}$ at higher orders}
\label{three}

It is straightforward to continue the analysis at three loops. As
expected, the cancellation of quartic and triple poles at three loops
in \eq{finms} is achieved automatically as a consequence of
lower-order constraints.  Double poles specify the relationship
between $\kappa_\psi$ and $\gamma_K$ at the three-loop level; using
\eq{simpol2} one can write
\beq
 \kappa_{\psi,0}^{(3)} = \gamma_K^{(3)} + \frac{b_0}{4} \kappa_{\psi,1}^{(2)}
 - \frac{b_0^2}{16} \kappa_{\psi,2}^{(1)} + b_1 \left(\kappa_{\psi,1}^{(1)}
 + \frac{3}{4} g_{U,0}^{(1)} \right)~.
\label{kappa3}
\eeq
As before, running coupling effects do not affect the known
relationship between $A(\alpha_s)$ and $\gamma_K(\alpha_s)$: demanding
the cancellation of of ${\cal D}_0(N)/\e$ terms at this order in fact
yields $A^{(3)} = \gamma_K^{(3)}/2$. $N$-independent single-pole
terms, on the other hand, yield the constraint
\beqa
 g_{U,0}^{(3)} + \frac{\kappa_{\psi,1}^{(3)}}{3} & = & - 4 B_\delta^{(3)} +
 2 \gtil^{(3)} + \frac{b_0}{4} \left(g_{U,1}^{(2)} + \frac{5}{6}
 \kappa_{\psi,2}^{(2)} \right) \nonumber \\ 
 & - & \frac{b_0^2}{16} \left(g_{U,2}^{(1)} + 
 \frac{11}{6} \kappa_{\psi,3}^{(1)} \right) + \frac{b_1}{4}
 \left(g_{U,1}^{(1)} + \frac{4}{3} \kappa_{\psi,2}^{(1)}\right)~.
\label{simpol3}
\eeqa
The finite coefficients of ${\cal D}_i (N)$ with $i = 1,2,3$ provide
nontrivial tests of the results achieved so far. Further,
concentrating on single logarithms, and using \eq{simpol3}, one finds
that
\beqa
 D^{(3)} & = & 4 B_\delta^{(3)} - 2 \gtil^{(3)} - \frac{b_0}{4}
 \left( g_{U,1}^{(2)} + \frac{\kappa_{\psi,2}^{(2)}}{2} \right) 
 + \frac{b_0^2}{16} \left(g_{U,2}^{(1)} + 
 \frac{3}{2} \kappa_{\psi,3}^{(1)} \right)
 - \frac{b_1}{4} \left(g_{U,1}^{(1)} + \kappa_{\psi,2}^{(1)}\right)
 \nonumber \\ & = &
 4 B_\delta^{(3)} - 2 \gtil^{(3)} - b_0 F_{\msb}^{(2)} - 
 \frac{b_1}{2} F_{\msb}^{(1)}~.
\label{D3}
\eeqa
The detailed structure of the coefficients in terms of the functions
$g_U$ and $\kappa_\psi$, as before, turns out to be irrelevant, and
the aswer is simply expressed in terms of lower order contributions to
the function $F_{\msb} (\alpha_s)$. This is remarkable, but easily
understood: in fact the details of the factorization given in
\eq{om2}, while conceptually crucial to prove formally the
exponentiation of logarithms to all orders, cannot affect the overall
structure of IR-collinear poles: one could, for example, define a
modified quark density including eikonal effects, and poles would
still cancel. Inspection of Eqs.~(\ref{D1}), (\ref{D2}) and (\ref{D3})
leads us then to the following all-order ansatz for the function
$D(\alpha_s)$, which summarizes the results of our work.
\beq
 D(\alpha_s) = 4 \, B_\delta (\alpha_s) - 2 \, \gtil (\alpha_s) 
 + \hat{\b} (\alpha_s) \, \frac{d}{d \alpha_s} F_{\msb} (\alpha_s)~.
\label{D} 
\eeq
The function $D(\alpha_s)$, governing threshold resummation for
electroweak annihilation at the single-logarithmic level, is thus
completely determined at order $\alpha_s^n$ by the knowledge of
virtual contributions to the nonsinglet splitting function, and
IR-collinear poles of the quark form factor, to the same order, plus
the value of exponentiated $N$-independent terms arising from real
emission at order $\alpha_s^{n - 1}$.
 
We are now in a position to evaluate the three-loop contribution to
the function $D(\alpha_s)$, thanks to the recent results of MVV.  The
three-loop contribution to the function $B_\delta (\alpha_s)$, in
fact, is given in Ref.~\cite{Moch:2004pa}; the three-loop coefficient
of the function $\gtil (\alpha_s)$ is given (in~\cite{Eynck:2003fn})
by the expression
\beq
 \gtil^{(3)} = G^{(3)}_0 - \frac{b_0}{4} G^{(2)}_1 - \frac{b_1}{4} G^{(1)}_1 
 + \frac{b_0^2}{16} G^{(1)}_2~,
\label{tilG3}
\eeq
and all relevant coefficients in the expansion of the function
$G(\alpha_s, \e)$ can be found in Ref.~\cite{Moch:2005id}, where MVV
use their results for DIS structure functions to evaluate explicitly
the quark form factor at three loops; finally, the value of $F_{\msb}
(\alpha_s)$ at two loops can be extracted by comparing our
exponentiated expression with the two-loop calculation of
Ref.~\cite{Hamberg:1991np}. We find
\beqa
 F_{\msbns}^{(2)} & = & \left( \frac{607}{324} - \frac{469}{144} 
 \zeta (2) + \frac{1}{4} \zeta^2 (2) - \frac{187}{72} \zeta (3) 
 \right) C_A C_F \nonumber \\ 
 & + &
 \left( - \frac{41}{162} + \frac{35}{72} \zeta (2) + 
 \frac{17}{36} \zeta (3) \right) n_f C_F~.
\label{FMSbar2}
\eeqa
Collecting all ingredients, or result for $D^{(3)}$ is
\beqa
 D^{(3)} \hspace{-8pt} & = & \hspace{-8pt} \left(- \frac{297029}{23328} + 
 \frac{6139}{324} \zeta (2) - 
 \frac{187}{60} \zeta^2 (2) + \frac{2509}{108} \zeta (3) -
 \frac{11}{6} \zeta (2) \zeta(3) - 6 \zeta(5) \right) C_A^2 C_F 
 \nonumber \\ & + & \hspace{-8pt}
 \left(\frac{31313}{11664} - \frac{1837}{324} \zeta (2) + \frac{23}{30}
 \zeta^2 (2) - \frac{155}{36} \zeta (3) \right) n_f C_A C_F 
 \nonumber \\ & + & \hspace{-8pt}
 \left(\frac{1711}{864} - \frac{1}{2} \zeta (2) -  \frac{1}{5} 
 \zeta^2 (2) - \frac{19}{18} \zeta (3) \right) n_f C_F^2 
 \nonumber \\ & + & \hspace{-8pt}
 \left(- \frac{58}{729} + \frac{10}{27} \zeta (2) + \frac{5}{27}
 \zeta (3) \right) n_f^2 C_F~.
\label{outD3}
\eeqa
The coefficient of the highest power of $n_f$ in $D^{(3)}$ can be
independently checked by comparing it with the renormalon calculations
of \cite{Beneke:1995pq} and~\cite{Gardi:2001di}: indeed, their results
agree with the last line of \eq{outD3}\footnote{We thank Einan Gardi
for pointing out this check to us and providing us with his results.}.

\sect{Discussion}
\label{discu}

We have analyzed threshold resummation for the Drell-Yan process in
the $\msb$ scheme, in light of the recent results obtained for Deep
Inelastic Scattering by MVV. Building upon a factorization proposed in
Ref.~\cite{Eynck:2003fn}, we have been able to derive a general
relationship expressing the function $D(\alpha_s)$, responsible for
threshold logarithms in the Drell-Yan cross section at
single-logarithmic level, in terms of data requiring the knowledge of
the virtual part of the nonsinglet splitting function, and the
singular terms in the quark form factor, at the same perturbative
order, plus a well-defined set of $N$-independent terms arising in the
Drell-Yan cross section at lower orders. Our main result is \eq{D},
and, using MVV results, it has enabled us to evaluate the three-loop
coefficient $D^{(3)}$, given in \eq{outD3}.

An immediate question is whether our results extend to the case in
which the hard annihilating partons are gluons, which is relevant for
the process of Higgs production via gluon fusion, in the effective
theory with the top quark integrated out. It is, in fact, easy to show
that an equation identical in form to \eq{D} holds also for
gluon-initiated electroweak annihilation, provided the various
functions involved are appropriately redefined: in fact, threshold
resummation in that case can still be cast in the form of \eq{finms},
with $2 A(\alpha_s)$ replaced by the cusp anomalous dimension for a
Wilson line in the adjoint representation, $2 A_g(\alpha_s)$, and two
new functions $D_g(\alpha_s)$ and $F_{\msb}^{g} (\alpha_s)$. The
$\msb$ distribution can be similarly defined for initial gluons, with
$B_\delta (\alpha_s)$ replaced by the virtual part of the appropriate
gluon splitting function.  The gluon form factor obeys an
exponentiation identical in form to \eq{mygam}. All ingredients are
thus in place to yield \eq{D}. A more delicate question is whether
this implies a simple relationship between the perturbative
coefficients of $D$ and $D_g$.  Up to two loops, one verifies by
explicit calculation~\cite{Harlander:2002wh,Ravindran:2004mb} that
$D_g$ can be obtained from $D$ by simply replacing the overall factor
of $C_F$ with $C_A$, just as one does in deriving $A_g$ from $A$.  It
is unlikely, however, that such a simple behavior will persist to all
orders: in fact, while it is natural to expect that purely eikonal
quantities such as $A$ or the function $U_R$ will be sensitive only to
the representation of the gauge group in which the eikonal line is
placed, not all information encoded in \eq{D} arises from eikonal
lines; it is known, for example~\cite{Moch:2005tm}, that subleading
poles in the gluon form factor cannot be obtained from the quark form
factor with such a simple prescription. Even eikonal functions would
probably require a more careful treatment at high enough order, when
high-rank Casimir operators constructed out of the symmetric $SU(N)$
tensors $d_{abc}$ come into play.

All this notwithstanding, we argue that at the three-loop level the
simple prescription is still valid, and one can in fact compute
$D_g^{(3)}$ by simply replacing the overall factor of $C_F$ with
$C_A$. To see it, one can make use of an observation of
Ref.~\cite{Ravindran:2004mb}, already exploited in
Ref.~\cite{Moch:2005tm}.  According to this observation, it is
possible to isolate in the quark form factor, and specifically in the
function $G(\alpha_s,\ e)$, a class of maximally nonabelian
contributions, dubbed $f_n^{(q,g)}$ in Ref.~\cite{Moch:2005tm}, which
exhibit the same behavior as the eikonal anomalous dimenson $A$ ({\it
i.e.} they obey the simple replacement rule, as verified up to three
loops in~\cite{Moch:2005tm}). We have explicitly checked up to three
loops that in fact the leading terms of our equation, $4
B_\delta^{(k)} - 2 \gtil^{(k)}$, coincide with the maximally
nonabelian factors $f_k^q$ up to an irrelevant multiplicative factor.
Since the remaining term in our \eq{D} is a running coupling effect,
determined at lower orders where the replacement rule is known to
apply, we conclude that indeed $D_g^{(3)}$ is also given by
\eq{outD3}, with the overall $C_F$ replaced by $C_A$.

We conclude by noting that we expect these results to be useful for
hadron collider phenomenology. In fact, along the lines of
\cite{Moch:2005ba}, the knowledge of $D^{(3)}$ allows to perform ${\rm
N}^3{\rm LL}$ threshold resummation for Drell-Yan and Higgs
production, to what is expected to be a very good approximation. This
can be used not only to provide a more accurate QCD prediction for
these processes, but also to check for the stability and the
convergence properties of both ordinary perturbation theory and the
expansion of its resummed counterpart in towers of
logarithms. Finally, we note that several of the building blocks of
our analysis also enter in resummations and high-order perturbative
calculations for more complicated processes at hadron colliders (see
for example~\cite{Sterman:2002qn}). It would be interesting to study
the extent to which our techniques can be applied also in that
context.

\vspace{1cm}

{\large {\bf {Acknowledgements}}}

\vspace{2mm}

\noindent We thank Lance Dixon for a stimulating discussion which
contributed motivation to perform this analysis. L.M. thanks Einan
Gardi for several discussions and for the test performed in
\secn{three}, as well as the CERN PH Department, TH Unit for
hospitality during the completion of this work, which was also
supported in part by MIUR under contract $2004021808\_009$.  The work
of E.L. is supported by the Foundation for Fundamental Research of
Matter (FOM) and the National Organization for Scientific Research
(NWO).

\vspace{1cm}

{\large {\bf {Note added}}}

\vspace{2mm}

\noindent While our paper was being written, S.~Moch and A.~Vogt
completed their own calculation of $D^{(3)}$, for both quark- and
gluon-initiated scattering~\cite{Moch:2005ky}, using a different line
of argument. Their results completely agree with ours.

\end{document}